# Efficient Instrumentation for Performance Profiling

Edu Metz, Raimondas Lencevicius
*Nokia Research Center*
*5 Wayside Road, Burlington, MA 01803, USA*
*Edu.Metz@nokia.com    Raimondas.Lencevicius@nokia.com*

## 1. Introduction

Performance profiling consists of tracing a software system during execution and then analyzing the obtained traces. However, traces themselves affect the performance of the system distorting its execution [5]. Therefore, there is a need to minimize the effect of the tracing on the underlying system's performance. To achieve this, the trace set needs to be optimized according to the performance profiling problem being solved. Our position is that such minimization can be achieved only by adding the software trace design and implementation to the overall software development process. In such a process, the performance analyst supplies the knowledge of performance measurement requirements, while the software developer supplies the knowledge of the software. Both of these are needed for an optimal trace placement. The following sections expand on this position.

## 2. Performance profiling

Performance profiling is the means of determining where a software system spends its execution time. It uses trace instrumentation to gather event data. Various types of event information can be obtained with traces, such as component entry and exit, function calls, software execution states, message communication, resource usage, etc. However trace instrumentation comes at a cost — it impacts the performance of a software system [3][6]. For example, resource tracing is most of the time more intrusive then tracing function calls.

Not only does event tracing take some time, adding traces changes the behavior of the software system because of additional memory and I/O accesses [1]. In addition, in a real-time software system, instrumentation could possibly result in violation of real-time constraints and timing requirements. Trace instrumentation reduces the validity of performance profiling, so instrumentation has to be kept to a minimum.

### 2.1. Minimizing Performance Impact

There is a need to minimize the performance impact of trace instrumentation. To achieve this, we need to create efficient instrumentation. To instrument effectively, it is essential to know what events to monitor during execution of the software system and what information to collect when the event occurs. When instrumenting the software, it is essential to understand the purpose and goals of each trace and how it will affect the instrumented software component. From the performance profiling point of view, a "good" trace not only records the required event information; it also minimizes the impact on the system's performance, and does not violate any constraints and requirements.

In choosing the instrumentation granularity, it is important to address the trade-off between the amount of event information required and the performance impact of the trace instrumentation. For example: permanent OS traces in the scheduler report when a task switch occurs. These traces do not indicate if the task switch is due to preemption by a higher priority task or completion of the current running task. The duration of a task activity cannot be calculated based on OS scheduling traces only. It requires additional instrumentation. However, these additional traces will further impact the performance of the software system.

It should be noted that creating an efficient instrumentation does not eliminate the performance impact of trace instrumentation but rather tries to minimize the performance impact.

Let us summarize what we just talked about: efficient instrumentation for performance profiling imposes the following requirements:
- minimize the number of instrumentation points
- minimize the runtime overhead, and
- guarantee constraints and requirements.

### 2.2. Efficient Instrumentation

We need to establish instrumentation that meets the requirements outlined in the previous section. This can be a complicated task, particularly in industry, where software development and performance profiling are often performed by different individuals each with their own set of skills and knowledge. Software developers have detailed knowledge of the software implementation. They understand the purpose of each instrumentation point and are able to assess the impact the instrumentation will have on the functional behavior of a software component. However, developers lack the understanding of what event data is needed. In addition, they may not be eager to insert event traces simply because they will not use them. On the other hand, performance analysts know what events need to be traced and understand what information needs to be recorded when an event occurs. However, performance analysts lack a detailed understanding of the software. We propose to draw upon the knowledge and skills software developers and performance analysts bring with them and use this knowledge to create efficient trace instrumentation.

To achieve this, we need to add trace instrumentation for performance profiling to the software development process. During the requirements phase the performance analyst should identify system-level performance requirements such as response time, throughput, and resource utilization, and start determining the events that need to be traced to check these requirements. For example, if the system level performance requirements state a maximum response time then the software's main entry and exit events (events e1 and e2 in Figure 1) need to be traced. However, it is not always possible to identify instrumentation points for all system level performance requirements during the requirements phase. For example, validation of resource utilization requirements requires knowledge of the software's execution states, which are not known until the design phase. Furthermore, only system level performance requirements are known during the specification phase. During the design phase, the performance analyst should identify lower level performance requirements such as messaging latency, interrupt response times, real-time deadlines, and time spent in the kernel. Next, the performance analyst should determine the events that need to be traced to check these requirements (for example, events e3 and e4 in Figure 1 as well as other events marked with black dots) and specify the event data that needs to be recorded when the event occurs. Typical events that need to be traced include: component entry and exit points, function calls, state transitions, message send and receive, and resource accesses. The developer then incorporates all the instrumentation requirements into the software design by identifying the corresponding instrumentation points. During implementation, the developer inserts traces at each event point, both manually and by activating (a subset of) permanent traces. The developer should plan to incrementally introduce the traces through iterations to minimize the impact of the instrumentation code on software system operations. During this process, the performance analyst should provide guidance to the software developer on choosing the instrumentation granularity (e.g., trace events e5 and e8, but not events e6 and e7 in Figure 1).

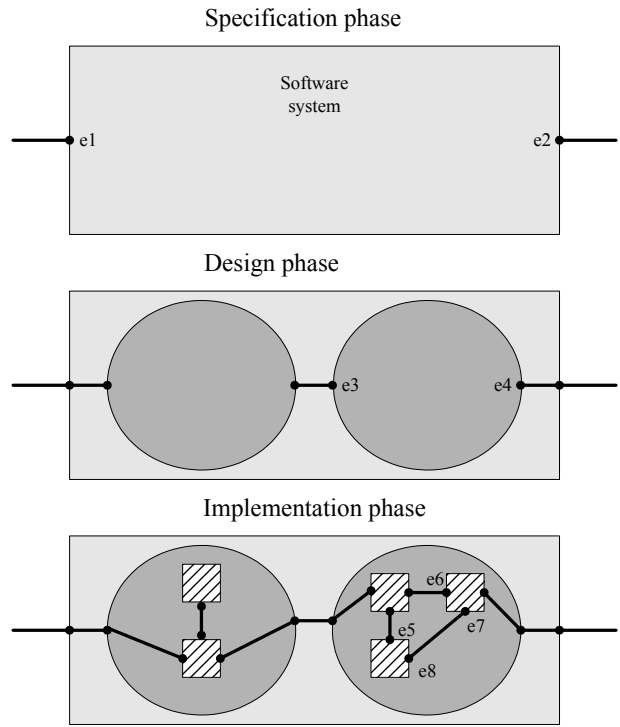

Figure 1: Trace design and implementation process

To illustrate this approach, let us look at an example. In mobile devices, power consumption is an important performance requirement [2]. The power consumption varies depending on the hardware resources used. During execution the software accesses hardware resources. These accesses need to be monitored to determine when a hardware resource is used, but should all access events be traced or is it enough to just trace enable and disable events? This question is best answered by the performance analyst. During the requirements phase the performance analyst identifies the power consumption requirements of the hardware resource. At design time, the performance analyst identifies the hardware access events that need to be traced to check the power consumption requirements. When tracing hardware access events in a mobile device it is very easy to violate real-time constraints and timing requirements. In addition, driver software of each hardware resource is unique. Instrumenting hardware drivers requires a detailed understanding of the software, and the developer is best suited for this task. During the design and implementation phase the developer incorporates the instrumentation requirements set by the performance analyst into the driver software.

A good follow through by both the performance analyst and software developer is essential for the success of the proposed approach. For example: during the actual performance profiling phase, the performance analyst should relay any kind of trace instrumentation inefficiencies to the developer. The developer in turn should make the necessary instrumentation improvements and provide the performance analyst with an updated instrumented software build in a timely manner.

The approach to adding trace instrumentation for performance profiling to the software development process addresses the requirements outlined in section 2.1. In addition, this approach would yield some other incentives:

- allows for creating built in 'standardized' performance trace instrumentation, and
- provides formatting rules for performance event data.

Smith and Williams [4] proposes a systematic approach to software performance engineering. They focus on estimating the performance of a software system during each stage of the software development process. Our approach attempts to optimize the performance impact of trace instrumentation for performance profiling by adding the software trace design and implementation to the overall software development process.

## 3. Summary

In this position paper, we described an approach to optimize trace instrumentation for performance profiling. The approach involves adding trace instrumentation for performance profiling to the software development process. It draws upon the knowledge and skills software developers and performance analysts bring with them — using this knowledge to create efficient trace instrumentation.

The proposed approach has the potential to decrease the number of instrumentation points. It would yield sufficient traces to profile the performance, yet it would not trace more event data than needed. In addition, the proposed approach would reduce the impact of trace instrumentation on software system performance.